\documentclass{aa}
\usepackage{graphicx}
\def\cm1{$\rm cm^{-1}$}
\def\kms{$\rm km\,s^{-1}$}
\def\\I{{\sc i}}
\usepackage{txfonts}
\begin{document}
   \title{The narrow, inner CO ring around the
magnetic Herbig Ae star, HD 101412}


   \author{Charles R. Cowley
          \inst{1}
          \and
          S. Hubrig\inst{2}\fnmsep\thanks{Based on observations
          obtained at the European Southern Observatory (ESO
          programme 087.C-0124(A)).}
          \and 
          F. Castelli\inst{3}
          \and 
          B. Wolff\inst{4}
          }

   \offprints{C. Cowley}

   \institute{Department of Astronomy, University of Michigan,
              Ann Arbor, MI 48109-1090, USA\\
              \email{cowley@umich.edu}
         \and
              Leibniz-Institut f{\"u}r Astrophysik Potsdam (AIP), An der Sternwarte 16, 
              D-14482 Potsdam, Germany \\
         \and
              Istituto Nazionale di Astrofisica, Osservatorio Astronomico di Trieste, Via Tiepolo 11, 
                 I-34143 Trieste, Italy \\
          \and
             European Southern Observatory, Karl-Schwarzschild-Str. 2, 85748,
             Garching bei München, Germany.
             }

   \date{Received ; accepted 23/12/211}

   \abstract{We describe and model emission lines in the
first overtone band of CO in the magnetic Herbig Ae star 
HD 101412.  High-resolution CRIRES spectra reveal unusually
sharp features which suggest the emission is formed in a
thin disk centered at 1 AU with a width 0.32 AU {\it or less}.
A wider disk will not fit the observations.
Previous observations have
reached similar conclusions, but the crispness of
the new material brings the emitting region into sharp focus.

            }
\titlerunning{The CO ring of HD 101412}
\authorrunning{Cowley, Castelli, Hubrig, Wolff}

\maketitle
%

\section{Introduction}
\label{se:intro}
The star HD 101412 (V1052 Cen) has been referred to in 
numerous papers as a Herbig Ae star.  It has an infrared
excess, and a disk (cf. Fedele, et al. 2008).  

HD 101412 is most unusual in having
resolved, magnetically split spectral lines which reveal a 
surface field modulus that varies between 2.5 to 3.5 kG
(Hubrig, et al. 2010). 
Salyk, et al. (2011a)\, have surveyed molecular emission in a 
variety of young stellar objects.  They found 
the emission to be much more subdued in Herbig Ae/Be stars
than their cooler congeners, the T Tauri stars.  This
was true for HD 101412 as well, which was among the 25
Herbig Ae/Be stars they discussed.  One exception, however,
was the molecule $\rm CO_2$, which had a very large flux
in HD 101412; indeed, only one T Tauri star had a higher
$\rm CO_2$ flux.

In this letter, we report observations of lines in the
first overtone (v=2 $\rightarrow$ 0)
band of carbon monoxide (CO) in HD 101412.  This feature
has been the subject of previous studies (cf. Najita 1996,
and references therein), leading to conclusions in general
agreement with those presented here.

Wheelwright, et al. (2010) studied this band in 7 upper
main-sequence stars with masses from 6 to 
43 $M_\odot$, significantly larger than that of HD 101412,
ca.  2.5 $M_\odot$ (Hubrig, et al. 2009). 
Individual lines were not resolved, but
models were constructed to fit the band head.
The fits constrained the CO emission
within limits of a few tenths of an AU (inner) to up to 8
AU (outer).  

\section{Observations and reductions}

We present observations made with the ESO CRIRES spectrometer
(K\"{a}ufl, et al. 2004) using a slit width of 0.2 arc seconds.
The achieved resolution  was determined from weak telluric
lines, and is slightly above 90000.
The signal to
noise is difficult to obtain empirically because of strong 
and myriad weak telluric features.  Local tests of the 
rms fluctuations in 1 to 3\, \AA\, regions 
( 10 to 30 pixels) yield S/N between 100
and 200, averaging 153.

We discuss observations in two CRIRES regions, 2290-2301nm,
and 2303-2313.5\,nm, made on 5 April 2011 at rotational phase 0.94
(Hubrig, et at. 2011).
The first of these regions contains
the band head of $\rm ^{12}C^{16}O$ (henceforth, CO will
mean this isotopologue).
This head is located at 2292.8976 nm, the position of the
rotational line R(51): 
($v'' = 0, J'' = 51 \rightarrow v' = 2, J'= 52$).
[Note: The double prime refers to the lower level.]

Stellar (rest) wavelengths  were 
determined using telluric and identified atomic
stellar features, primarily C \\I, Mg \\I and Si \\I lines.  These
lines were measured in other CRIRES regions, covering
1065--1105 and 1585--1550 nm.  The transformation to
rest stellar wavelengths was then applied to the CO regions.

\begin{figure}
   \centering
   \includegraphics[width=0.34\textwidth,angle=270]{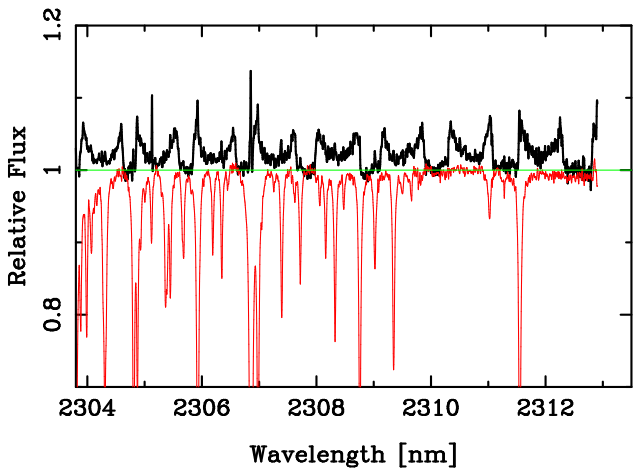}
      \caption{The black spectrum shows CO emission lines 
        R(20)(right) to R(27)(left) in HD 101412 after removal of telluric
        lines using the standard star,
        HR 4537, j Cen. (red).}
         \label{fig:one}
   \end{figure}

Telluric features were removed using hot
standard stars with large rotational velocities, observed
at comparable zenith distances.  For the CO features, this
procedure introduced both noise, and artifacts; it was
necessary to judge the position of the continuum in both
the target and standard stars.  The minimum of the central
M-feature is particularly sensitive to this subjective
judgment (see Fig.\ref{fig:one}).

%
%
\section{The CO emission lines and band head}
\subsection{Optically thick or thin?}

We assume an optically thin emitting gas as a reasonable
approximation.  This is based on 
the sharpness of the profiles shown in Figs.~\ref{fig:one}
and \ref{fig:two}.  Self-absorption would smooth the sharp
edges of the M-shaped profiles, filling in the centers, and
spilling intensity over at the sides (edges) of the profiles.
Additionally, we obtain a good fit to the band head using
this assumption (Fig.~\ref{fig:three}).  
We can also make an estimate of the optical depth from
the excess CO emission above the stellar continuum, using
stellar parameters from Cowley et al. (2010)
and Hubrig et al. (2009).  The method
is similar to that discussed by Salyk et al. (2011b), though
we use a toroidal geometry for the emitting region rather
than a slab.  Plausible assumptions lead to an optical depth
of 0.04 for R(20).  Details may be found at 
\newline http://www.astro.lsa.umich.edu/~cowley/oldindex.html

\subsection{An empirical fit}

All observations discussed here are in the R-branch of the
first overtone band of the ($X^2\Sigma^+$) ground electronic 
state of CO.  
Identifications of molecular features, and temperature-dependent
predictions are based on the data file coxx.asc 
(Kurucz 1993).  Only $\rm ^{12}C^{16}O$ lines were considered.
All others were assumed to be too weak to be relevant, or
outside our limited observational regions.

   \begin{figure}
   \centering
   \includegraphics[width=0.34\textwidth,angle=270]{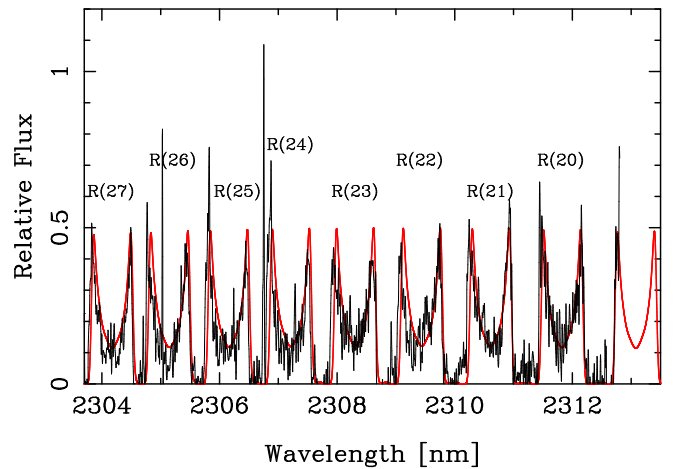}
      \caption{Observed CO R(20)-R(27) in black.  The observations
    are those of Fig.~\ref{fig:one} (black).  We subtracted 
    unity, and scaled the remainder vertically to fit the
    observations.  R-branch labels
    are written above the M-shaped profiles.  Rest wavelengths
    of these lines correspond to the inner vertex of the M's.
    R(19) is
    at the extreme right, but only the violet edge of the observed
    profile is seen.  Several sharp spikes may be attributed to
    imperfect removal of the telluric features that may be
    seen in the standard star spectrum of Fig.~\ref{fig:one}.
    Red shows calculated
    molecular features for a temperature of 2500K. 
         \label{fig:two}}
   \end{figure}
%
%
 
\begin{figure}
   \centering
   \includegraphics[width=0.34\textwidth,angle=270]{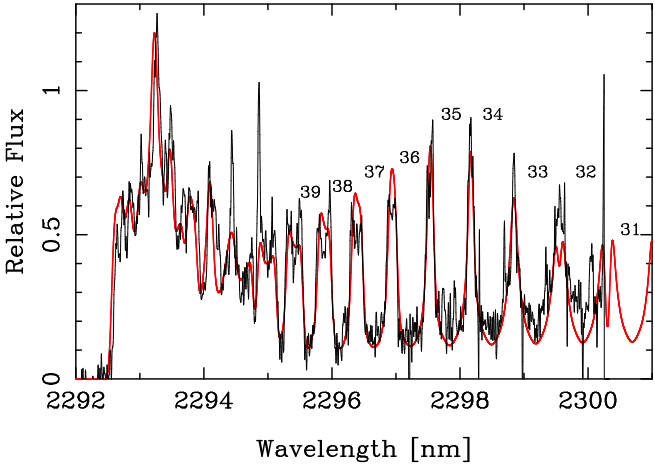}
      \caption{First overtone band head with partially cosmetic
    fit.  The assumed splitting and relative intensities 
    (within a given M-shaped line) of the
    multiple Gaussian profiles are the same as in 
    Fig.~\ref{fig:two}.  The overall distribution of intensities
    is somewhat better fit by assuming $T$ = 2500 K rather 
    than 2000 or 3000 K.}
         \label{fig:three}
   \end{figure}

Emission spectra (Figs.~\ref{fig:two} and ~\ref{fig:three}) 
were calculated with the help of Kurucz's data. 
We assumed that emission flux profiles can be represented by
gaussian functions with central intensity 
$I\propto gf\lambda^{-3}\exp{-\chi_u/kT}$, where gf and 
$\chi_u$ are the oscillator strength and the excitation potential 
of the upper
level of the CO transitions. We adopted T=2500\,K.
 
In many cases (see Section ~\ref{sec:model}), one can determine 
a rotational temperature from the relative intensities of
the individual lines (cf. Brittain, Najita, and Carr 2009).
However, the relative intensities of
R(20)-R(27) hardly change from $T = 2000$ to 3000 K, the 
range we believe appropriate (see below).  
A first calculation, using a single Gaussian profile for each line
placed the emissions in the middle of the observed
M-shaped profiles.   
Carr (2005) discusses the canonical, disk-model 
M-shaped profile.
We eventually adopted a set of 21 Gaussians for each line, with
0.0666 nm FWHM (0.04 nm 1/e-width).  They were
weighted appropriately to reproduce 
the observed profiles (red, Fig.~\ref{fig:two}).
The Gaussian at the center, therefore had the lowest weight.
Other choices for the FWHM and number of Gaussians could
give similar results.
The fit is thus cosmetic,
apart from the central wavelength matches, and the 
overall profile widths
of the calculated features.  These match very well the molecular 
wavelengths and the  expected widths
(see Section~\ref{sec:model}).

The velocity half-widths of the Gaussians, for
a mean wavelength of 2309 nm is 47.8 km s$^{-1}$.  This
value, uncertain by   $\approx$ 2\%,
is in agreement with the Keplerian velocity 
47.1 \kms used below in a disk-model fit to the profiles.


The strongest lines in the region of Fig.~\ref{fig:two} are
R(20)-R(27).  However, higher-order lines, with wavelengths
increasing for R(n $>$ 51) fill in some of the region between 
the stronger features.  These R(n $>$ 51) lines diminish in
intensity for increasing values of n.

Fig.~\ref{fig:three} shows the region of the band head.  
Values of $J''$ are written over the centers of the M-shaped
profiles.  Because
of convergence to the head, vertical ``sides'' of the
profiles approach one another, coincide, and eventually cross.  
Additionally, close to the head, transitions in
the R-branch leading ``away'' (to the red) from the head (e.g. R(52)..R(56)) 
overlap {\it and} are of comparable intensity to the lines
with R-values below that of the head, R(51).  

The badly-blended head is reasonably well
fit by the multi-Gaussian model.  This indicates that the 
assumed profiles remain a good approximation for rotational
lines somewhat higher than R(39).  However, the relative
importance of these higher-series lines diminishes rapidly.
The R(51) line is 3.3 times weaker than R(31) at 2500 K.

The overall shape of the head region is more sensitive to the
rotational temperature than the region of Fig.~\ref{fig:two}.
By-eye fits to the observed features were best for temperatures
between 2000 and 3000\,K.  The plot is for 2500\,K, with an
uncertainty of several hundred degrees.

\section{A model emission ring\label{sec:model}}

In this section we discuss a physical model to reproduce
the M-shaped profiles.  The code is far simpler than 
those used in some cited studies (e.g. Salyk 2011b).
The simplification is possible because we assume (1) the
emitting gas is optically thin, (2) in pure Keplerian rotation,
and (3) unobscured by the star or a dust disk.

The relevant code calculates the
profile of a single emission line, and has not been adapted
for multiple emissions, such as those shown in Figs.~\ref{fig:two}
and ~\ref{fig:three}.

We constructed a two-dimensional model of the emission 
region, consisting of a set of 81 concentric rings,
each with 288 emitting gas elements.  
They are all in Keplerian rotation around
a 2.5 $M_\odot$ star.  The rings may be placed at 
arbitrary radial distances from the star. 
We assume that each element emits a
Gaussian profile with a FWHM of 0.0666\,nm.

A model
that marginally fits the observations, has rings 
centered at 1 AU, spaced by 0.004 AU.  This
gives a disk 0.32 AU wide.  A wider disk, at 1 AU, 
gives a profile
that is too broad and rounded at the peaks to fit
the observations.   Indeed
a single ring would produce an 
entirely satisfactory fit.

According
to Fedele et al. (2008)  the disk plane is $10^\circ$
from the line of sight.  Since $\cos(10^\circ) = 0.985$, we
ignore the inclination, apart from the assumption that it
must be enough for the emitting ring to be visible.
A cosine factor may easily be incorporated, should future
work reveal a
larger inclination to the line of sight.
Contributions from the 81 x 288 elements are
Doppler shifted, added linearly, and  scaled vertically to fit the 
observations.  

\begin{figure}
   \centering
   \includegraphics[width=0.34\textwidth,angle=270]{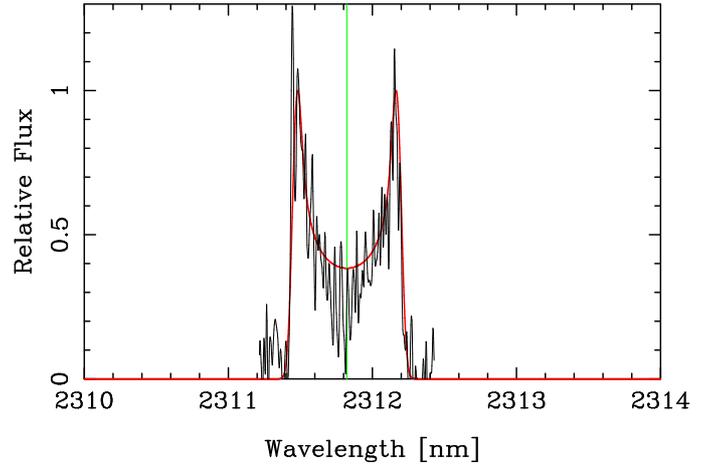}
      \caption{Model fit (red) to the R(20) profile 
      shown in Fig.~\ref{fig:two}.  A thin, vertical line (green)
      marks the rest wavelength of R(20).
      The model is virtually a
      single ring, since $r_{\rm in} = 0.96$ and 
      $r_{\rm out} = 1.04$ AU.}
         \label{fig:four}
   \end{figure}

Fig.~\ref{fig:four} shows a fit of the ring model to 
the rotational line R(20).  The  81 rings are centered at
1 AU, and are separated by 0.001 AU.

\begin{figure}
   \centering
   \includegraphics[width=0.34\textwidth,angle=270]{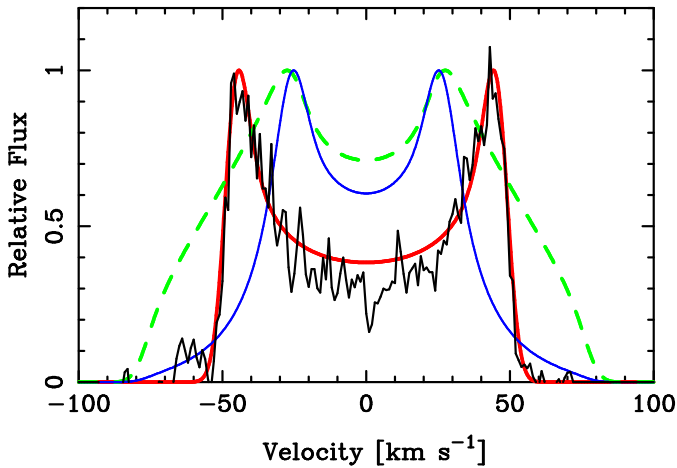}
      \caption{Model fit to a composite of R(20) through R(27)
       of Fig.~\ref{fig:two}.  The same parameters used for
       the fit of Fig.~\ref{fig:four} (red)
       give excellent agreement for the composite.  Two
       additional curves illustrate profiles that would result
       from a wider disk (see text).}
         \label{fig:five}
   \end{figure}

Salyk et al. (2011b) stacked a number of their profiles in
order to increase the signal to noise.  We have attempted this
in Fig.~\ref{fig:five}.  The abscissa is rendered in velocity
space.  Plots have been scaled vertically for an optimum
fit.  The overall shape of the composite profile (black) is well
fit by the model, apart from the low regions near the
center of the `M' (thick red).
It is unclear how significant this is.
We noted earlier the sensitivity of this portion of the
profile to subjective judgments of the continuum location.

Additional emission is seen at the foot of the violet 
(or negative-$V$) side
of the M-shaped profile.  This is surely due to R-branch 
lines, R($J"$), where $J" > 51$.  There is no
provision in the model for this effect.
%

Fig.~\ref{fig:five} shows two additional curves which 
illustrate the effect of an extended disk.  Both curves
assumed a disk with innermost rim at 0.4 AU, and outermost
rim at 3.6 AU.  The solid blue curve
was made assuming each ring had the same total emissivity.
The dashed green curve shows a
profile where the emission is weighted by the negative
1.5$^{\rm th}$ power of the radial distance.  In this
case, the innermost rings, with highest Keplerian velocities,
have higher weights.  This causes the additional emission 
at larger positive and negative velocities.  

It is clear that such a disk cannot possibly account for 
sharply-edged rotational profiles.  Quite elaborate radial
intensity variations were considered by Acke,
et al. (2005, cf. their Fig.\,21),
in connection with [O I] emission.  Happily, such complexity 
is unnecessary in the present case.


\section{Discussion}

Van der Plas et al. (2008), and Fedele et al. (2008) have
discussed the protoplanetary disks around ``three  young
intermediate mass stars,'' including HD 101412, based on
observations of forbidden oxygen, [O \\I] $\lambda$6300.  
They discuss disk models, though not as localized
as the one that fits our CO observations.  

Salyk, et al. (2011b) studied the fundamental
(v=1 $\rightarrow$ 0) CO band in 31 T Tauri and Herbig Ae/Be
stars, including a number with transitional disks.  Rotational
lines of the fundamental CO band are more commonly seen in
T Tauri stars than the first overtone, the subject of the
present paper.  
Salyk et al. conclude their line fluxes 
are ``consistent with emission from a single temperature 
ring width 0.15 and 0.01 AU respectively...''.  This conclusion
is very close to that of the present paper.  However, the
profiles illustrated by Salyk, et al. are much less sharply defined
than those we observe in the first CO overtone in HD 101412.  
The central portions of their profiles are mostly filled in,
indicating emission from regions of a disk beyond the innermost
rim.  Thus, they were also able to
obtain fits (see their Table 4) with disk models extending over
several and even tens of AU.  The HD 101412 observations are
incompatible with emission from such an extended disk.

Perhaps the T Tauri star with CO (fundamental) profiles most 
closely resembling the first overtone lines of HD 101412 is 
V836 Tauri (Najita, Crockett\& Carr 2008).  
But note the sloping sides
of their averaged profiles (their Figs. 4 and 5) 
in comparison to the
nearly vertical sides of an analogous profile in HD 101412 
presented below (Section ~\ref{sec:model}).


The existence of narrowly-confined emission regions
is well established.  The current observations may be
the best-defined example found thus far.

\begin{acknowledgements}

CRC thanks his Michigan colleagues for advice and support.
Special thanks to Lee Hartmann for advice and 
encouragement.  We acknowledge with thanks, the comments
of Eric Mamajek.

\end{acknowledgements}

\end{document}